\documentclass[5p,twocolumn,showpacs,nofootinbib,showkeys,superscriptaddress,preprint]{elsarticle}

\usepackage{lineno,hyperref}
\modulolinenumbers[5]
\usepackage{graphicx,floatflt}
\usepackage{amsmath}
\usepackage{amstext}
\usepackage{amssymb}
\usepackage{epsfig}
\usepackage[usenames]{color}
\usepackage[dvipsnames]{xcolor}
\newcommand{\dd}{\mathrm{d}} % For the derivatives
\newcommand{\rr}{\mathrm}

\newcommand{\rhogw}{\rho_{\rm gw} }
\newcommand{\rhoc}{\rho_{\rm c} }

\newcommand{\Msun}{M_\odot}
\newcommand{\Mc}{{\mathcal M}_{\rr c}}

\newcommand{\OM}{\Omega_{\rm M}}
\newcommand{\del}{\delta_{\rm PBH}^{\rm local}}

\newcommand{\mPBH}{m_{\rm PBH}}

\def\gtrsim{\mathrel{\hbox{\rlap{\hbox{\lower4pt\hbox{$\sim$}}}\hbox{$>$}}}}

\def\lesssim{\mathrel{\hbox{\rlap{\hbox{\lower4pt\hbox{$\sim$}}}\hbox{$<$}}}}

\newcommand{\be}{\begin{equation}}
\newcommand{\ee}{\end{equation}}
\newcommand{\ba}{\begin{eqnarray}}
\newcommand{\ea}{\end{eqnarray}}

\begin{document}

%\preprint{TTK-16-42, IFT-UAM/CSIC-16-098}

\title{Detecting the gravitational wave background 
from primordial black hole dark matter}

%\author{S\'ebastien Clesse}
%\email{clesse@physik.rwth-aachen.de}
%\affiliation{Institute for Theoretical Particle Physics and Cosmology (TTK), RWTH Aachen University, D-52056 Aachen, Germany}

%\author{Juan Garc\'ia-Bellido} \email{juan.garciabellido@uam.es} 
%\affiliation{Instituto de F\'isica Te\'orica UAM-CSIC, Universidad
%Auton\'oma de Madrid, Cantoblanco, 28049 Madrid, Spain}

\author{S\'ebastien Clesse}
\ead{clesse@physik.rwth-aachen.de}
\address{Institute for Theoretical Particle Physics and Cosmology (TTK), RWTH Aachen University, D-52056 Aachen, Germany}

\author{Juan Garc\'ia-Bellido} \ead{juan.garciabellido@uam.es} 
\address{Instituto de F\'isica Te\'orica UAM-CSIC, Universidad
Auton\'oma de Madrid, Cantoblanco, 28049 Madrid, Spain}

\date{\today}

\begin{abstract}
The black hole merging rates inferred after the gravitational-wave detections by Advanced LIGO/VIRGO and the relatively high mass of the progenitors are consistent with  models of dark matter made of massive primordial black holes (PBH).    PBH binaries emit gravitational waves in a broad range of frequencies that will be probed by future space interferometers (LISA) and pulsar timing arrays (PTA).    The amplitude of the stochastic gravitational-wave background expected for PBH dark matter is calculated taking into account various effects such as initial eccentricity of binaries, PBH velocities, mass distribution and clustering.   It allows a detection by the LISA space interferometer, and possibly by the PTA of the SKA radio-telescope.   Interestingly, one can distinguish this background from the one of non-primordial massive binaries through a specific frequency dependence, resulting from the maximal impact parameter of binaries formed by PBH capture, depending on the PBH velocity distribution and their clustering properties.  Moreover, we find that the gravitational wave spectrum is boosted by the width of PBH mass distribution, compared with that of the monochromatic spectrum.  The current PTA constraints already rule out broad-mass PBH models covering more than six decades of masses, {but evading the microlensing and CMB constraints because black holes appear spatially distributed in clusters}.   
\end{abstract}
%\pacs{98.80.Cq}
\maketitle

\section{Introduction}

The recent detection by Advanced LIGO of the gravitational waves (GW) emitted by the coalescence of black hole binaries~\cite{Abbott:2016blz,Abbott:2016nmj} and the unexpected high mass of the involved BH (36, 29, 14 and 8 solar masses $\Msun$)  have revived the interest for models of primordial black holes (PBH) that could constitute a large fraction or even the totality of the Dark Matter (DM)~\cite{Clesse:2015wea,Bird:2016dcv,Clesse:2016vqa,Sasaki:2016jop,Carr:2016drx}.   But do the present astrophysical constraints allow such a population of massive black holes?  Do the detected BHs have a primordial origin~\cite{Raccanelli:2016cud,Raccanelli:2016fmc,Kovetz:2016kpi}?  If so, is it possible that these PBH account for the totality of DM?  The debate has been re-opened by the AdvLIGO discovery.  

On the \textit{pro-}side, it has been shown that consistent merging rates with the one inferred by AdvLIGO can be obtained for two simple PBH-DM models, one extrapolating the DM halo mass function towards small scales~\cite{Bird:2016dcv}, and one where PBH are regrouped in dense sub-halos such as the ultra-faint dwarf galaxies~\cite{Clesse:2016vqa}.  The latter model would also provide a natural  explanation to the \textit{missing satellite} and \textit{too-big-to-fail} problems if the existence of thousands of such ultra-faint satellite galaxies were confirmed, in which the PBH population could prevent the formation of stars and make their trajectory unstable in such an environment.    Furthermore it has been shown that PBH-DM  halos could explain some unexpected fluctuations in the near-IR cosmic infrared background (CIB), found to be coherent with the unresolved soft X-ray background~\cite{Kashlinsky:2016sdv}.  

On the \textit{con-}side, it has been proposed that BHs as massive as the ones detected by AdvLIGO could also result from a particular stellar evolution in low-metallicity environments~\cite{Belczynski:2016obo}.   
{Moreover a population of massive PBH accounting for the totality of DM could have induced signatures in the CMB anisotropy angular power spectrum\footnote{CMB spectral distortions are also expected and Ref.~\cite{Ricotti:2007au} claimed that masses larger than $\Msun$ are ruled out by FIRAS.  However Eq.~(44) in~\cite{Ricotti:2007au} has a factor $(1+z)^{-2}$ missing, which leads to much less stringent constraints (by a factor of a few millions).  There is now an agreement in the community that PBH-DM is not constrained by FIRAS.} ~\cite{Ricotti:2007au,Chen:2016pud}, or should have been detected through microlensing events of stars in the Magellanic clouds~\cite{Tisserand:2006zx,Alcock:1998fx,Griest:2013aaa}, although it is debated and model dependent~\cite{Hawkins:2015uja,Sasaki:2016jop}.   But the physical processes leading to signatures in the CMB are subject to large uncertainties.  Especially the accuracy of the Bondi accretion approximation is not well established and could be suppressed by the large BH velocities in the case of early clustering.   }  {Recently it has been claimed that the crucial mass window between $10$ and $100 M_\odot$ was closed by the observations of the central star cluster within the Eridanius II dwarf galaxy~\cite{Brandt:2016aco,Green:2016xgy}.  However, the stability of the stellar cluster in Eridanus II seems to suggest that there is an IMBH at its center, which would soften these bounds~\cite{Li:2016utv} and make a broad mass spectrum model compatible with observations.}

%Moreover the role that could have initially high merging rates in increasing the mean PBH mass is still unclear and could be a natural way to evade distortion constraints.  
Finally, regarding the microlensing constraints from EROS, MACHOS and Kepler, they are naturally evaded if the PBH are clustered in the galactic halo so that the probability of finding such a cluster in the line-of-sight of the Magellanic clouds were the observations were done is actually very low~\cite{Clesse:2016vqa}.   {With a similar argument, one can evade constraints on masses larger than $100 M_\odot$ from the non-disruption of wide binaries, because the probability of disruption is suppressed by the probability of encounter with a PBH cluster.  }

As a consequence, further observations will be required to distinguish between the different models and in order to validate or definitively rule out the massive PBH-DM hypothesis.   Future observations of numerous merging events, possibly involving even more massive BHs, or binaries with high orbital eccentricities, would allow to distinguish between a stellar or a primordial origin~\cite{Cholis:2016kqi,Kovetz:2016kpi}.   They could also allow to reconstruct the PBH mass spectrum and reveal how clustered PBH are~\cite{Clesse:2016vqa}, which would help to reveal their mechanism of formation in the early Universe~\cite{GarciaBellido:1996qt,Garcia-Bellido:2016dkw}.   The next runs of LIGO/VIRGO could increase dramatically the number of detected PBH mergers.   Another rich source of information could come from the detection of (or from the constraints on) a stochastic background of gravitational waves induced by a huge population of massive PBH~\cite{1999PhRvD..60h3512I,Cholis:2016xvo,Mandic:2016lcn,Schutz:2016khr}.  

{The goal of this paper is to compute the expected stochastic background of GWs produced by massive PBH clustered in compact halos and with some mass distribution, as in Refs.~\cite{Clesse:2016vqa,Clesse:2015wea,GarciaBellido:1996qt,Garcia-Bellido:2016dkw,Blinnikov:2016bxu,Garriga:2015fdk}.  We examine more particularly how the GW background produced by primordial BH binaries could be distinguished from the one of BH stellar binaries, due to the limited impact factor in the process of PBH capture that affects the GW spectrum at frequencies probed by future space interferometers and pulsar timing arrays (PTA).   Our analysis also considers the effects of eccentricities and velocity distributions, as well as the case PBH have a broad mass spectrum, instead of having all the same mass as in the recent analysis of~\cite{Bird:2016dcv,Sasaki:2016jop,Cholis:2016kqi,Cholis:2016xvo,Schutz:2016khr}.}   Furthermore we discuss the detectability of the signal, not only with future Earth-based GW detectors like KAGRA~\cite{Somiya:2011np} and ET\footnote{http://www.et-gw.eu/}~\cite{Luck:2009zz,Punturo:2010zz}, but also with future GW detectors in space such as LISA\footnote{http://sci.esa.int/lisa/}~\cite{Heinzel:2014ets,Bartolo:2016ami}, DECIGO~\cite{Kawamura:2006up} and BBO~\cite{Harry:2006fi}, as well as with current and future limits from pulsar timing arrays such as EPTA\footnote{http://www.epta.eu.org}~\cite{Lentati:2015qwp,Lazarus:2016dlu}, IPTA~\cite{vanHaasteren:2013vf} and SKA\footnote{https://www.skatelescope.org}~\cite{Smits:2008cf,Zhao:2013bba}.   

Our main result is that the GW background induced by PBH-DM models consistent with AdvLIGO rates is detectable by LISA, and possibly by the SKA-PTA if their mass distribution is sufficiently broad.   Indeed a broad mass spectrum is found to strongly boost the stochastic GW amplitude, so that the current PTA already constrain broad-mass PBH-DM models with $10^{-2} \Msun \lesssim \mPBH \lesssim 10^4 \Msun $.   Moreover, we find that the GW spectrum exhibits a very specific frequency dependence that will help to distinguish it from the one of non-primordial BH binaries.   This effect is directly linked to the PBH clustering, enhancing their velocities and reducing the maximal impact parameter in the PBH capture process.  This implies a minimal frequency for the GW emitted by binaries formed through PBH capture.  Typically, for velocities larger than tens of km/s, e.g. corresponding to PBH clustering within ultra-faint dwarf satellite galaxies, we predict a suppressed spectrum at PTA frequencies and a slight but detectable modification w.r.t. the typical $ f^{-2/3} $ frequency dependence of the strain $h_c(f)$, on frequencies probed by LISA.  

The paper is organized as follows:  The formalism used to calculate stochastic backgrounds of gravitational waves is summarized in Section~\ref{sec:GWcalc}.  In Section~\ref{sec:MC} we describe our synthetic population of PBH binaries.  The effect of initial eccentricity is calculated in Section~\ref{sec:eccentricity}.  Section~\ref{sec:merging} focuses on PBH merging rates in both the monochromatic approximation and more realistic broad mass spectrum case.   In Section~\ref{sec:GWbkg} we compute the stochastic GW background for our model and discuss its detectability with future GW experiments and PTAs.  Our results are summarized and some perspectives are presented in Section~\ref{sec:ccl}.  

\section{Stochastic backgrounds of gravitational waves}  \label{sec:GWcalc}

A population of massive BH binaries {experiencing merging} induces a stochastic background of gravitational waves (GW), that is characterized by the relative GW energy density to the critical density $\rho_{\rr{c}}$ today, per unit interval of logarithmic frequency, 
\be\label{OmGW}
\Omega_{\rr{gw}} (f) \equiv \frac{1}{\rho_{\rr c} } \frac{\dd \rhogw(f)}{\rr d \ln f}~,
\ee
where $f$ is the observed GW frequency.   Another key quantity is the GW strain amplitude $h_{\rr c} $, given by~\cite{Phinney:2001di,Sesana:2008mz}
\be
h_{\rr c}^2 (f) = \frac{4 G }{\pi f^2}  \frac{\dd \rhogw(f)}{\rr d \ln f}.
\ee
The GW energy spectrum is given by the superposition of the GW radiation coming from merging BH binaries over the whole cosmic history,
\be \label{rhogw}
 \frac{\dd \rhogw(f)}{\rr d \ln f} = \int_0^\infty   \frac{\dd z}{1+z} \ \frac{\dd n}{\dd z} \ \frac{\dd E_{\rr{gw}}}{c^2 \dd \ln f_{\rr r}},
\ee
where $\dd E_{\rr{gw}}  /  \dd \ln f_{\rr r} $ is the released energy by the merging events, per logarithmic interval of the rest-frame frequency $f_r = f(1+z)$.  In the Newtonian limit, it is well approximated by 
\ba \label{eq:rhogw}
\frac{\dd E_{\rr{gw}}}{c^2 \dd \ln f_{\rr r}} &=& \frac{\pi^{2/3} }{3 c^2} \Mc^{5/3} (G f_{\rr r})^{2/3} F(e) \, \\[2mm]
F(e) &=& \Big(1-e^2\Big)^{-7/2}\!\left(1 + \frac{73}{24}e^2 + \frac{37}{96}e^4\right)\,,
\ea
within the range $f_{\rr{min}} < f_r < f_{\rr{ISCO}}$, where $e$ is the eccentricity of the orbit and $\Mc$ is the chirp mass of the BH binary, defined as $\Mc^{5/3} = m_{\rm A} m_{\rm B} M^{-1/3}$, with  $m_{\rm A}$ and $m_{\rm B}$ the two BH masses and $M=M_{\rm tot} = m_{\rm A} + m_{\rm B}$.   The value of the minimal frequency $ f_{\rr{min}} $ is related to the initial BH separation $a_0$,   
\be
f_{\rm min} = \frac{(GM)^{1/2}}{\pi a_0^{3/2}}\,.
\ee
The maximal frequency $ f_{\rr{ISCO}} \approx 4.4\,{\rm kHz} \, (\Msun / M) $ corresponds to the innermost stable circular orbit.   For stellar-mass BHs, this maximal frequency is well above the frequencies probed by PTAs and space interferometers, but it is within the range of  AdvLIGO and other earth-based interferometers, for typical BH of several tens of solar masses.     {As explained later, the typical initial BH separation can range from $10^{-5}$ to $10^3$ astronomical units (AU) for BH regrouped in $10^6 - 10^9 \Msun$ halos, corresponding to frequencies ranging from $100$ down to $10^{-11} \, {\rm Hz}$.    }

In the scenario considered here, BH binaries are formed when two PBH trajectories cross sufficiently close to each other so that they become bounded and start inspiralling, until they merge, which does not immediately produce circular orbits but instead highly eccentric orbits.    Soon after starting spiraling around each other, the emission of GW also radiates angular momentum and the eccentricity goes to zero much faster than energy emission~\cite{PhysRev.136.B1224}.  Nevertheless we analyse in Section~\ref{sec:eccentricity} for the low-frequency part of the GW spectrum, close to the initial emission, whether some eccentricity remains for more than one orbit, and whether or not it should be taken into account in the stochastic GW background calculation.

In Eq.~(\ref{eq:rhogw}), $\dd n / \dd z$ is the number density of GW events within the redshift interval $[z, z+\dd z]$, related to the merger rate $\tau^{\rr{merg}}$ in a comoving volume through
\be
\frac{\dd n}{\dd z} = \tau_{\rr{merg}} \frac{\dd t}{\dd z} = \frac{\tau_{\rr{merg}} }{(1+z) H(z)}~.
\ee
The Hubble rate is given by
\be
H^2(z)= H_0^2  \left[ (\Omega_{\rr{DM}} + \Omega_{\rr b})  (1+z)^3 + \Omega_{\rr r} (1+z)^4  + \Omega_\Lambda \right],
\ee
where $\Omega_{\rr {DM}}$, $\Omega_{\rr b} $, $\Omega_{\rr r}$ are respectively the today dark matter (DM), baryons (b), radiation (r) and cosmological constant ($\Lambda$) energy densities relative to the critical density.  Throughout the paper we assumed the best fit values after Planck~\cite{Ade:2015xua}. 

In the PBH-DM model of Ref.~\cite{Clesse:2016vqa}, the PBH are assumed to be clustered within compact virialized halos already at high redshifts, whose size and density are almost constant in time. This means that the merging rate per unit of comoving volume is approximately constant over the whole cosmic history.  Therefore, assuming a monochromatic PBH mass distribution, one gets a characteristic strain today
\ba  
h_{\rr c}^2 (f) & = & \frac{4\ \tau_{\rr{merg}}}{3 \pi^{1/3} c^2}   (G \Mc)^{5/3} \, F(e)\, f^{-4/3}  \nonumber \\ && \ \times  \int_0^{z_{\rr{max}}} \hspace{-3mm} \frac{\dd z }{H(z) (1+z)^{4/3}} \,, \label{eq:hc2_mono}
\ea
with $ \tau_{\rr{merg}} $ in units of $ \rr{yr}^{-1} \rr{Gpc}^{-3}$, which gives
\be \label{eq:hcf}
h_{\rr c} (f) \simeq 1.1 \times 10^{-25} \tau_{\rr{merg}}^{1/2}\!\left(\frac{f}{\rr{Hz}}\right)^{-2/3}\hspace{-1mm}\left( \frac{\Mc }{\Msun}\right)^{5/6}  \hspace{-2mm} F(e)^{1/2}\,.
\ee
This is the right order of magnitude {for a detection of the GW stochastic background produced by} inspiralling BH binaries of a few tens of solar masses in the range of AdvLIGO, for merger rates in the tens of events per year per Gpc$^3$.

{One goal of the paper is go beyond the approximation of a monochromatic mass distribution, which is not supported by the AdvLIGO events if they all involve BH of primordial origin.}.   The case of a broad mass distribution is considered in Sec.~\ref{sec:broadspectrum}.  In this case, the merging rate depends on the two progenitor masses $m_{\rm A} $ and $m_{\rm B}$.  One therefore has to integrate Eq.~(\ref{eq:hc2_mono})† over $m_{\rm A} $ and $m_{\rm B}$ in order to get the strain, 
\ba     \label{eq:hc2}
\hspace*{-3mm} && h_{\rr c}^2 (f)  =   \frac{4\,G^{5/3} f^{-4/3}}{3 \pi^{1/3} c^2}   \int_0^{z_{\rr{max}}} \frac{\dd z}{H(z) (1+z)^{4/3}}   \\
\hspace*{-3mm} && \times  \int\!\!\int \dd (\log_{} m_{\rm A}) \,\dd (\log_{10} m_{\rm B}) \,\tau_{\rm merg}(m_{\rm A},\,m_{\rm B}) \, \Mc^{5/3}\,,  \nonumber
\ea
where $\tau_{\rr{merg}}(m_{\rm A}, m_{\rm B})$ is now the merging rate per comoving volume, per unit of logarithmic mass interval.   { Eq.~(\ref{eq:hc2}) is valid for any model where the time-dependance of the merging rate is negligible.   It is straightforward to extend it to the case of a separable function of time and PBH masses, i.e.
$\tau_{\rr{merg}}(z,m_{\rm A}, m_{\rm B}) = t(z) \times \tau_{\rr{merg}}(m_{\rm A}, m_{\rm B})$.  Then the function $t(z) $ simply goes into the redshift integral, which only renormalizes the GW spectrum. }

\section{Synthetizing a population of PBH binaries} \label{sec:MC}

 \begin{figure}[h!]
    \begin{center}%\vspace*{-6mm}
        \includegraphics[width=6.9cm]{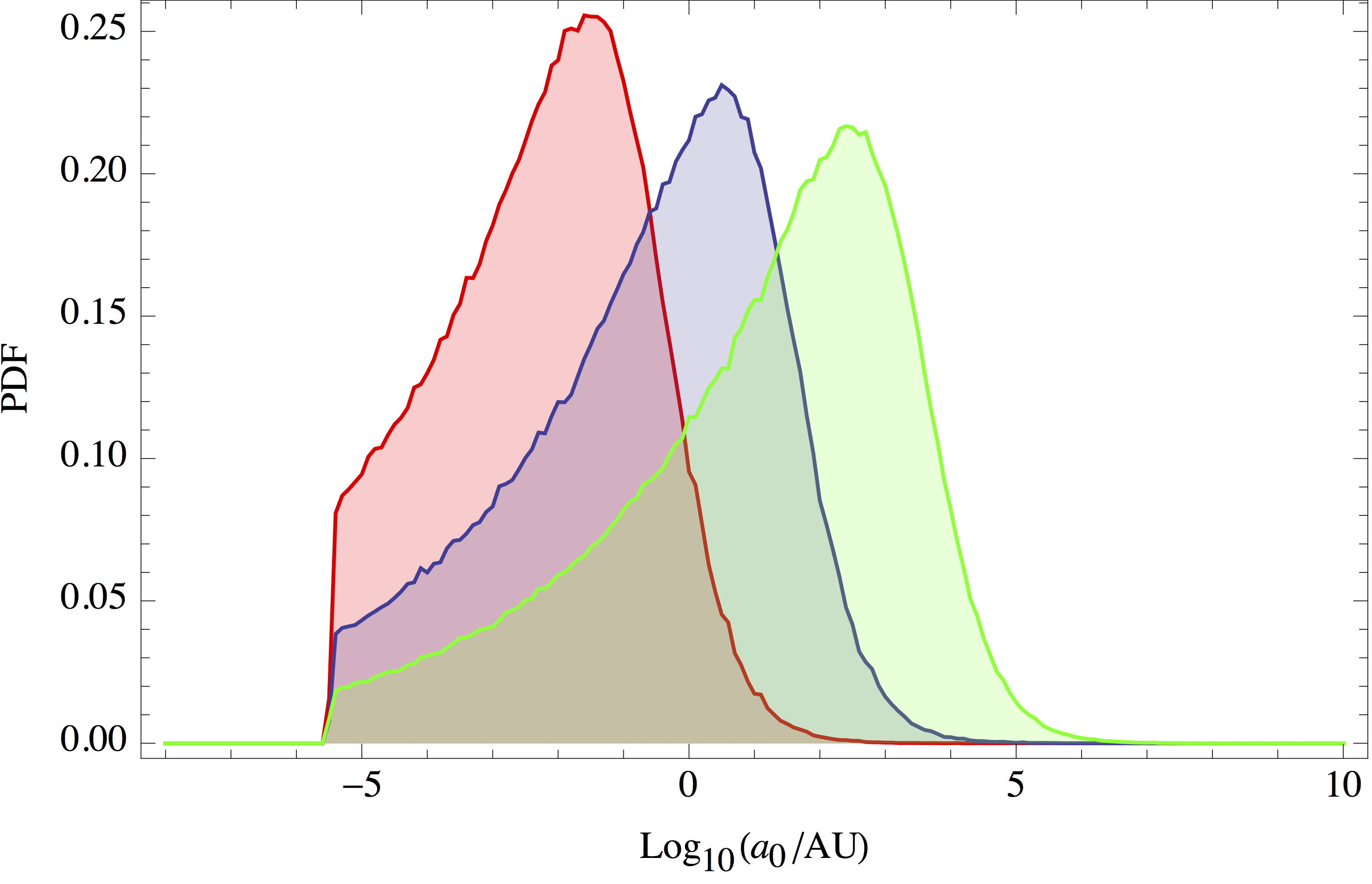} \\
        \includegraphics[width=6.9cm]{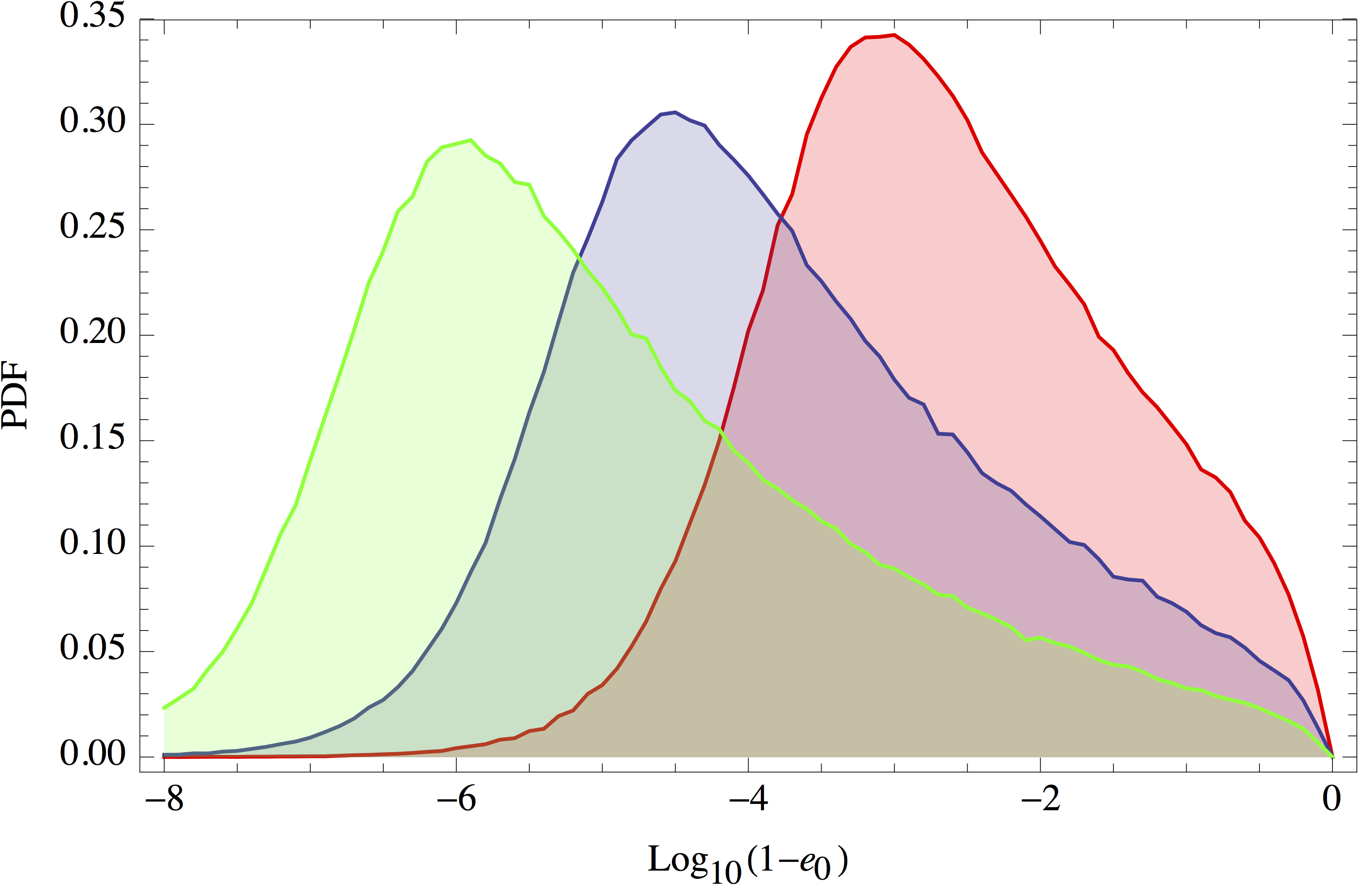}\\
        \includegraphics[width=6.9cm]{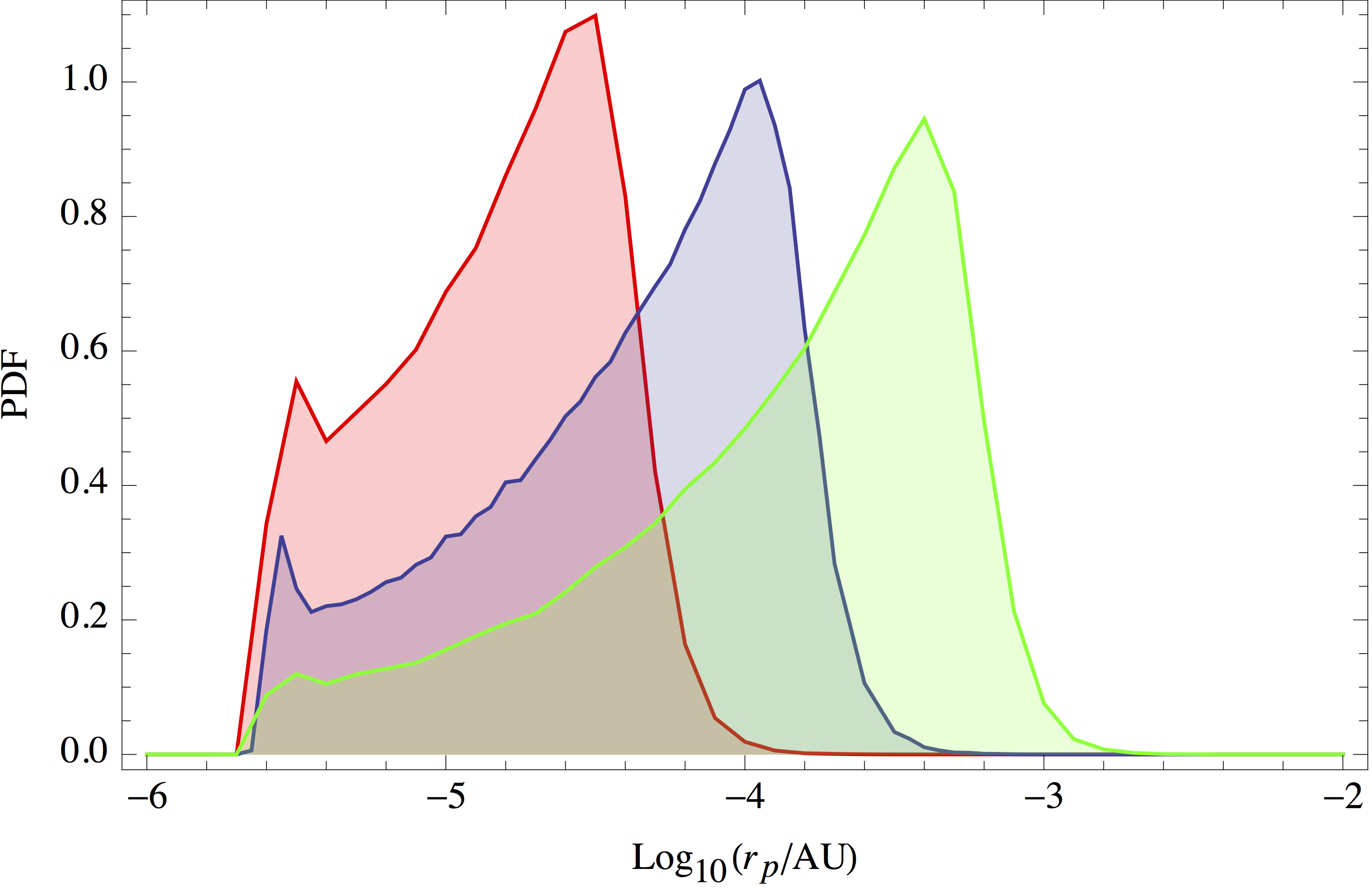}\\
         \includegraphics[width=6.9cm]{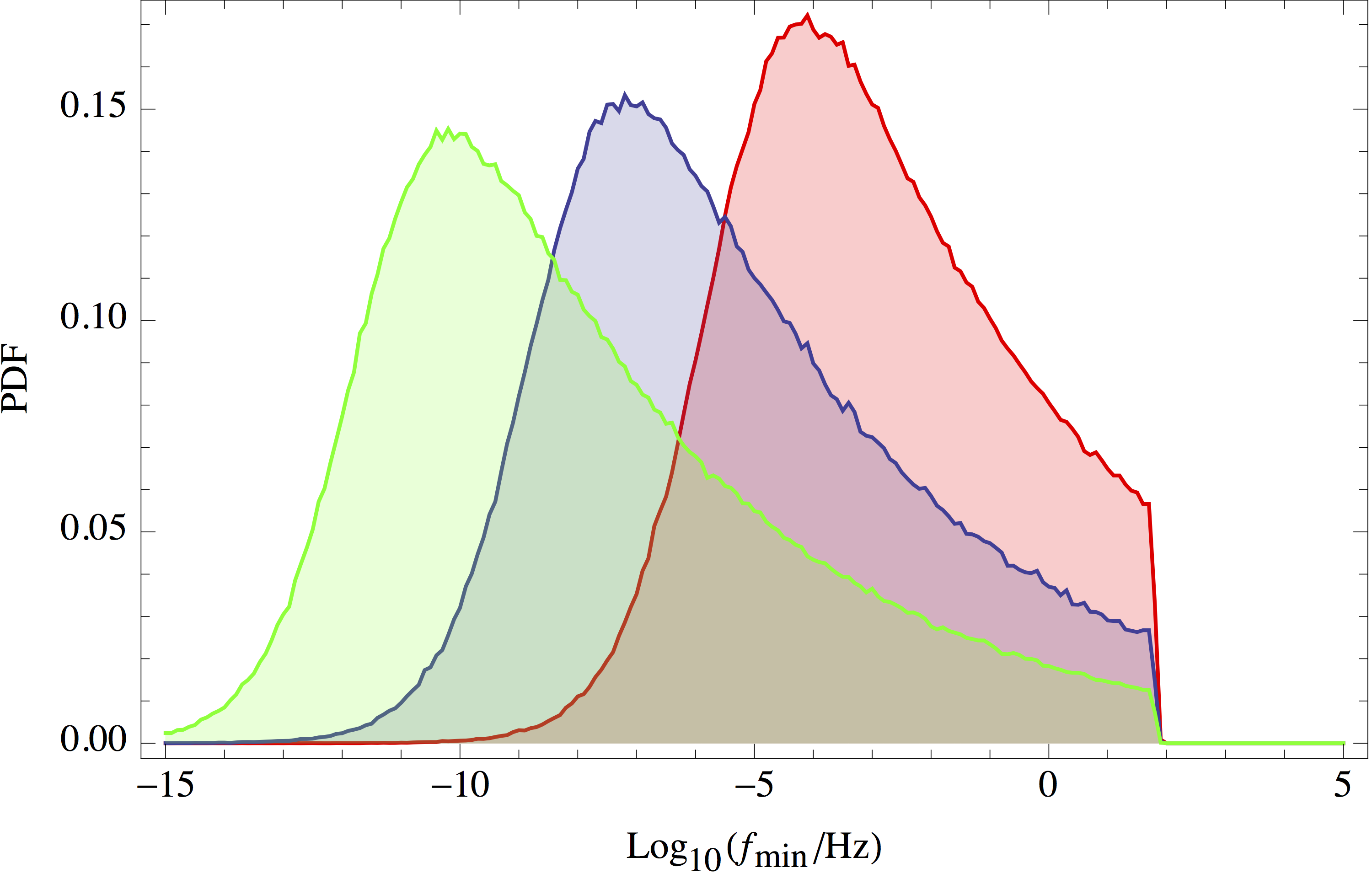}
%    \vspace*{-5mm}
    \end{center}
    \caption{PDF of initial orbit parameters $a_0$, $1-e_0$, $r_{\rr p}$ and $f_{\rm min}$ for a synthetized population of $10^6$ PBH binaries and halo Virial velocities $v_{\rm vir } = 2$ (green), $20$ (blue) and $200$ (red) km/s. } 
    \label{fig:PDFs}
\end{figure}

In order to cross-check our analytical calculations and to ease the inclusion of new effects such as PBH velocities, general mass distributions and cluster density profiles, we have synthesized a representative population of $10^6$ PBH-binaries, {at the time of their formation, given some relative velocity, impact parameter and mass distributions.   For each binary, we have then calculated the initial orbital parameters (eccentricity, semi-major axis, pericenter) and computed the expected GW strain amplitude, which has finally been averaged out over the whole population to get the final stochastic GW spectrum, marginalized over the above mentioned effects.}
% over all the considered effects is then computed from the posterior probability distribution of the strain amplitude.  

We have considered the case where all the PBH do not have the same mass but instead they  follow a broad, lognormal, distribution 
\be
\rho(\mPBH)= \frac{\delta_{\rr{PBH}}^{\rr{loc.}} \rho_{\rm DM}^0 }{\sqrt{2 \pi \sigma_{\rr{\rm PBH}}^2} } \exp \left[ - \frac{\log_{10}^2 (\mPBH / \mu_{\rr{\rm PBH}})}{2 \sigma_{\rr{\rm PBH}}^2} \right]~.
\ee
of central mass $\mu_{\rr{\rm PBH}}$ and  width $\sigma_{\rr{\rm PBH}}$ (note that logarithms are $\log_{10}$). 

We followed the picture detailed in \cite{Cholis:2016kqi,1998PhRvD..58f3003I}.  Binaries are formed when the energy radiated during the close encounter of two PBH, in the form of GW, is large enough for PBH to become gravitationally bounded to each other.  The released energy during this capture process is given by
\be
E_{\rm capt}   = \frac{85 \pi  \eta^2 (G M)^{9/2}}{12 \sqrt 2\, c^5 G \, r_{\rr p 0}^{7/2} },
\ee
where $\eta \equiv m_A m_B / (m_A + m_B)^2$ and $m_A, m_B$ are the two progenitor masses.    The pericenter $ r_{\rr p 0}$ is related to the initial orbit semi-major axis $a_0$ and eccentricity $e_0$ through $ r_{\rr p 0} = a_0 (1-e_0) $, themselves related to the initial PBH relative velocity $w$ and to the impact parameter $b$ through
\be  \label{eq:a0}
a_0 = G M w^{-2} \left[ \left(  \frac{b}{b_{\rm max} } \right)^{-7} - 1 \right]^{-1},
\ee
\be \label{eq:e0}
1- e_0^2 = \left( \frac{340 \pi \eta }{3}   \right)^{2/7} \left(\frac{w}{c}\right)^{10/7} \left(  \frac{b}{b_{\rm max}}\right)^2  \left[ \left(  \frac{b}{b_{\rm max} } \right)^{-7} - 1 \right],
\ee
{where $M \equiv m_A + m_B$.}
The maximal impact parameter reads
\be
b_{\rm max} = \left( \frac{340 \pi \eta }{3}   \right)^{1/7} \frac{G M}{c^2} \left(\frac{w}{c}\right)^{-9/7}.
\ee
We considered a uniform distribution of $b$ between $0$ and $b_{\rm max}$ but excluded too small values of $b$ leading to $e_0^2 < 0$.  This implies some cut-off in the distribution of $a_0$ and $f_{\rm min}$.  As in~\cite{Cholis:2016kqi} we assumed that $w = \sqrt 2 v_{\rm PBH} $.  We considered a Maxwellian distribution of PBH velocities,
\be
P(v_{\rm PBH}) = P_0 v_{\rm PBH}^2 \rr e^{-\frac{v_{\rm PBH}^2}{2 v_{\rm vir}^2}},
\ee
with $P_0$ a normalization factor and $v_{\rm vir}$ the Virial velocity expected for the PBH cluster.  

The PDFs of $1-e_0$, $a_0$ and the corresponding $f_{\rm min}$ of our synthesized binaries have been represented on Fig.~\ref{fig:PDFs} for $M = 60 M_\odot$ and $\sigma = 0.1$ and  $v_{\rm vir} = 2$, $20$ and $200 \, \rr{km/s}$.

\section{The effect of eccentricities}  \label{sec:eccentricity}

{Since the GW emission from inspiralling BH orbits also radiates away angular momentum, most BH binary systems contributing to the stochastic GW background have circularized their orbit.  Nevertheless one has to take account that a small fraction of them, the ones with orbits that are neither close to circular neither close to parabolic, i.e. $e_0 \sim \mathcal O(0.1)$ can affect the GW spectrum.  In particular, they can modify the $ f^{-2/3}$ behavior of the characteristic strain on frequencies associated to the first revolutions and thus could leave a signature of the primordial BH origin.  In order to determine how important is this effect, we must take into account the eccentricity of the orbits in Eq.~(\ref{eq:hc2_mono}).}

The evolution of the semimajor axis $a$ of the orbit as it loses angular momentum can be computed exactly as a function of the eccentricity~\cite{PhysRev.136.B1224},
\ba
a &=& a_0 \, G(e)/G(e_0)\,, \\
G(e) &=& \frac{e^{12/19}}{1-e^2} \left(1 + \frac{121}{304}\,e^2\right)^{870/2299}\,.
\ea
%where $\bar G(e)\equiv G(e)(1-e)$,
The initial semimajor axis $a_0$ and eccentricity of the orbit $e_0$ are related by the condition that the closest approach at periastron should not be smaller than the Schwarzschild radius of the system,
\be\label{emax}
r_{\rr p} = a_0(1-e_0) > R_S = 3\, {\rm km}\,\frac{M}{\Msun}\,.
\ee

It is now possible to write the eccentricity dependence of the GW emission in terms of these initial conditions,
\be\label{Fe}
F(e) = 1 + 2\left(\frac{a}{r_{\rr p}}\right)^{5/2}\hspace{-2mm}\bar G(e_0)^{5/2}
+ 3\left(\frac{a}{r_{\rr p}}\right)^{7/2}\hspace{-2mm}\bar G(e_0)^{7/2}\,,
\ee
where $\bar G(e)\equiv G(e)(1-e)$, and where the (restframe) semimajor axis depends now on frequency and mass,
\be
a = \frac{(GM)^{1/3}}{(\pi f_r)^{2/3}} =  \frac{(GM)^{1/3}}{(\pi f)^{2/3}} (1+z)^{-2/3}\,,
\ee
and thus $F(e)$ depends on mass, frequency and redshift. Integrating over redshift and marginalizing over $(r_{\rr p},\, e_0)$, we obtain the final stochastic spectrumof Eq.~(\ref{OmGW}). 
However, as it will be shown in Sec.~\ref{sec:mergertime}, the time it takes for the system to radiate its angular momentum and end up with no eccentricity, $e\simeq 0$, is less than a single orbit, and thus the effect of eccentricity can be safely neglected in the computation of the stochastic GW background.  
%The effect of eccentricity of the orbit induces a small correction to the usual GW spectrum, although more pronounced at lower frequencies, where PTA and space based GW antennas are sensitive. 

\subsection{Integration over redshift}

The contribution of eccentricity $F(e)$ in Eq.~(\ref{Fe}) to the GW amplitude~(\ref{eq:hc2_mono}) has three terms. The first one corresponds to the integral
\ba\nonumber
A_1&=&\hspace{-1mm}\int_0^{z_{\rr{max}}} \hspace{-3mm} \frac{\dd z }{H(z) (1+z)^{4/3}} = \frac{1}{H_0\sqrt{\OM}}\times \\[2mm]
&& \times \frac{6}{11}\,{}_2F_1\!
\left(\frac{1}{2},\frac{11}{18},\frac{29}{18},\frac{\OM-1}{\OM}\right)\,,
\ea
where $_2F_1 $ is the hypergeometric function. The second one corresponds to
\ba\nonumber
\hspace*{-2mm} A_2&=&\hspace{-1mm}\int_0^{z_{\rr{max}}} \hspace{-3mm} \frac{\dd z }{H(z) (1+z)^3} = \frac{\OM}{4\,H_0(1-\OM)^2}\times \\[2mm]
\hspace*{-2mm} &&\times\left[\frac{2-\OM}{\OM} - {}_2F_1\!
\left(1,\frac{-1}{3},\frac{1}{6},\frac{\OM-1}{\OM}\right)\right]\,.
\ea
And the third one,
\ba\nonumber
\hspace*{-2mm} A_3&=&\hspace{-1mm}\int_0^{z_{\rr{max}}} \hspace{-3mm} \frac{\dd z }{H(z) (1+z)^{11/3}} =  \frac{3\OM}{16\,H_0(1-\OM)^2}\times \\[2mm]
\hspace*{-2mm}  &\times & \left[\frac{2+5\OM}{\OM} - 7\,{}_2F_1\!
\left(1,\frac{-1}{9},\frac{7}{18},\frac{\OM-1}{\OM}\right)\right]\,.
\ea
All these terms become $(0.7642, 0.3602, 0.2945)\,H_0^{-1}$, for
$\OM=0.3$

\subsection{Merger time of inspiralling BH}  \label{sec:mergertime}

We can now assemble all the pieces together, but before we do that, we should compute the time it takes for the eccentricity to go to zero for initial values of $(M,\ a_0,\, e_0)$. This was computed by Peters in~\cite{PhysRev.136.B1224}, for $e_0\simeq1$
\ba\nonumber
t_{\rm decay} &=& 2.24\times10^{12}\,{\rm yrs}\left(\frac{a_0}{1\,{\rm AU}}\right)^{4}\left(\frac{\Msun}{M}\right)^{3}(1-e_0^2)^{7/2}\\ \label{tedecay}
&=& 1.96\times10^{13}\,{\rm yrs}\left(\frac{r_{\rr p}}{1\,{\rm AU}}\right)^{4}\left(\frac{\Msun}{M}\right)^{3}\,,
\ea
after marginalization over initial eccentricities, and writing the expression in terms of periastron values. If we now marginalize over periastron with its PDF, we get
\be
t_{e-{\rm loss}} = 0.715\,{\rm days}\left(\frac{\Msun}{M}\right)^{3}\,.
\ee
Therefore, for heavy BH masses, one finds eccentricity-loss times that are shorter than the orbital period. We will then have to assume in those cases that the GW emission occurs in circular orbits and thus the standard formula is valid. The GW emission timescale therefore sets up a lower bound on the frequency, which constrains the effect of the eccentricity on the spectrum.  

Alternatively, we can compare the time of decay of the eccentricity (\ref{tedecay}) with the period of the binary,
\be
T \simeq \pi\,{\rm yrs} \left(\frac{a_0}{1\,{\rm AU}}\right)^{3/2}\left(\frac{\Msun}{M}\right)^{1/2}\,.
\ee
Therefore, in order for the decay to occur in more than a single orbit one needs to satisfy
\be \label{eq:e0limit}
1-e_0^2 > 0.00041 \left(\frac{M(\Msun)}{a_0({\rm AU})}\right)^{5/7}\,,
\ee
which for $M=60~\Msun$ and $a_0=100~{\rm AU}$, one obtains $1-e_0 > 1.42\times10^{-4}$. 
%This number is much larger than those considered in the previous sections, so for all these cases one can ignore the effect of eccentricity on the GW background.
{In addition, by definition one has always $\eta \equiv m_A m_B /(m_A + m_B)^2 < 1/4$.   One can use Eqs.~(\ref{eq:e0}) and (\ref{eq:a0}) to rewrite the inequality~(\ref{eq:e0limit}) in the limit $b/b_{\rm max} \ll 1$.  This gives $\eta \gtrsim 4 \times 10^5$, independently of $M$, $w$ and $b$, which is therefore never satisfied.  Considering the limit  $b/b_{\rm max} \simeq 1$ does not help, because the r.h.s. of the inequality{~(\ref{eq:e0limit})} then blows up by a factor $(1-b)^{-12/7}$ and one would require even larger values of $\eta$ to satisfy it.} 
Therefore in the context of a standard two-body capture it is impossible that the binary preserves a large enough eccentricity over more than one orbit, independently of the impact parameter and the initial velocities.    This was confirmed numerically with our synthetized population of binaries, never satisfying the constraint of Eq.~(\ref{eq:e0limit}).  The effect of eccentricities can thus be safely neglected in the calculation of the stochastic GW background. 

It is nevertheless worth noticing that the rapid angular momentum loss implies that important energies are also released in the form of gravitational waves, in a time scale going from seconds to hours depending on the PBH mass.  This should induce strong bursts of GW at the moment of capture.   The study of these bursts and the question of their detectability are left for a future work.

\section{Primordial Black Hole  merger rates}  \label{sec:merging}

If PBH are uniformly distributed in the halo of massive galaxies, their cross section is so tiny that it is impossible to reproduce merging rates in the range inferred by AdvLIGO, going from 9 to 240 yr$^{-1}$Gpc$^{-3}$~\cite{TheLIGOScientific:2016pea}.   However if they are clustered within smaller sub-halos, their merging rates are boosted.   This case was considered in Ref.~\cite{Bird:2016dcv} by extrapolating towards small scales the observed DM halo mass function.  We focus here on the alternative model of Ref.~\cite{Clesse:2016vqa} where PBH are regrouped in virialized sub-halos such as ultra-faint dwarf galaxies.   This model was characterized by a single parameter:  the local density contrast $\delta_{\rr{PBH}}^{\rr{loc.}}$ of PBH inside such clusters, compared to the cosmological DM density.     Values  $\delta_{\rr{PBH}}^{\rr{loc.}} \sim 10^{9} -10^{10}$ are typical to the cold DM density in globular clusters and ultra-faint dwarf galaxies detected by Keck/DEIMOS~\cite{Martin:2007ic,Simon:2007dq} and DES~\cite{Drlica-Wagner:2015ufc} and induce compatible merger rates with the ones inferred by AdvLIGO.  

In this section, we summarize how to obtain the corresponding merging rates following~\cite{Clesse:2016vqa}, A) in the simplest, less realistic case of a monochromatic mass distribution, B) by extending the calculation to the broad spectrum case.   We also extend previous calculations by marginalizing the PBH distribution over some sub-halo density profile.  

\subsection{Monochromatic mass spectrum}

The PBH individual merging rate is given by the capture cross-section $\sigma^{\rr{capt}}$ of two black-holes in dense clusters, which has been calculated in Refs.~\cite{1989ApJ...343..725Q,Mouri:2002mc}.   The criteria for crossing BH to form a bound system is that the energy lost in the form of gravitational waves is of the order of the kinetic energy.   Once gravitationally bounded to each other, the two BH quickly merge, typically in less than a million years.   In the Newtonian approximation, this capture rate is given by~\cite{Mouri:2002mc}
\ba 
\tau_{\rr{\rm PBH}}^{\rr{capt}} & = & 2 \pi \ n_{\rm PBH}(m_{\rr A})  v_{\rm PBH} \  \left( \frac{85 \pi}{6 \sqrt 2} \right)^{2/7} \nonumber \\  
& & \times \frac{G^2 (m_{\rr A} + m_{\rr B})^{10/7}  (m_{\rr A} m_{\rr B})^{2/7}  c^{18/7}} {c^4 w^{18/7}}    \label{eq:capture_rate}
\ea
where $n_{\rr{\rm PBH}} $ is the PBH number density and $w = \sqrt{2} v_{\rm PBH}$ is the relative velocity of the two BHs\footnote{In~\cite{Clesse:2016vqa} a different assumption, $w = v_{\rm PBH}$ was considered, inducing a difference of $2^{9/7}$ in the merging rates.}.  Because the cross-section is much larger than the BH surface area, the Newtonian approximation is sufficiently accurate.   

One can then define the local density contrast $\del$ as  $ n_{\rr{\rm PBH}} \equiv \del  \times \bar \rho_{\rm DM} / m_{\rr{\rm PBH}} $, with $\bar \rho_{\rm DM} = \Omega_{\rm DM} \rho_{\rr{c}}$ the mean DM cosmological density.   {It is considered here as a model parameter describing the PBH clustering.}   Many complex phenomena (e.g. merging, gas accretion, dynamical friction\,\dots) should be taken into account to achieve a proper and precise understanding of the process of PBH clustering.   We are presently far from this situation and this parameterization is aiming at capturing phenomenologically this process.   The capture rate, referred now as the individual merging rate, in a comoving volume is then
\be
%\tau^{\rm merg}_{\rm indiv}  \simeq  4.9 \times 10^{-29} ~ f_{\rm DM} \del \left( \frac{\mPBH}{\Msun} \right) \rr{yr}^{-1}\, , 
\tau^{\rm merg}_{\rm indiv}  \simeq  4.1 \times 10^{-29}  \del \left( \frac{\mPBH}{\Msun} \right) \rr{yr}^{-1}\,
\ee
whereas the total rate per Gpc$^3$ comoving volume is obtained by multiplying it by the PBH number within this volume, 
\be  \label{eq:merging}
\tau^{\rr{merg}}  \simeq  2.9 \times 10^{-9} ~f_{\rm DM} \del \ \rr{yr}^{-1} \rr{Gpc}^{-3}.
\ee
The factor $f_{\rm DM}$ is the fraction of DM made of PBH inside those halos.  It is lower than one if a non negligible fraction of PBH also populate e.g.~the galactic halos and/or the extragalactic medium, or if PBH constitute only a fraction of the total DM.  It is worth noticing that the merging rate is independent of the PBH mass.  In the considered model, it is constant in time, which corresponds to the assumptions that PBH clusters are already virialized at high redshifts and do not evolve much in time.  

Finally, one notes that the merging rate within a single halo has a $n_{\rm PBH}^2$ dependence.  Thus one can marginalize merging rates over some specific halo density profile $\rho(r)$ simply by replacing $\del $ by $ \langle \del \rangle f_{\rm profile} $ with $f_{\rm profile} = \langle \rho^2 \rangle / \langle \rho \rangle^2 $, where averaging is done over the halo volume up to some radius above which the remaining PBH will produce negligible merging rates.  For instance, in the case of a common Einasto profile, for a total halo mass of $10^5 \Msun$ and an averaging radius $20 R_{\rr s} = 400 \rr{pc} $, one gets $\langle \del \rangle f_{\rm profile} \approx 5 \times 10^8$ with a corresponding Virial velocity of about $70$km/s. 

%It is possible to go beyond the approximation of constant density contrast by marginalizing over some halo density profile.  One then gets
%\be
%\tau^{\rr{merg}}  \simeq  4 \times 10^{-9} ~f_{\rm DM} \del \ \rr{yr}^{-1} \rr{Gpc}^{-3}.
%\ee

\subsection{Broad mass spectrum}  \label{sec:broadspectrum}

 \begin{figure}
    \begin{center}%\vspace*{-6mm}
    \includegraphics[scale=0.65]{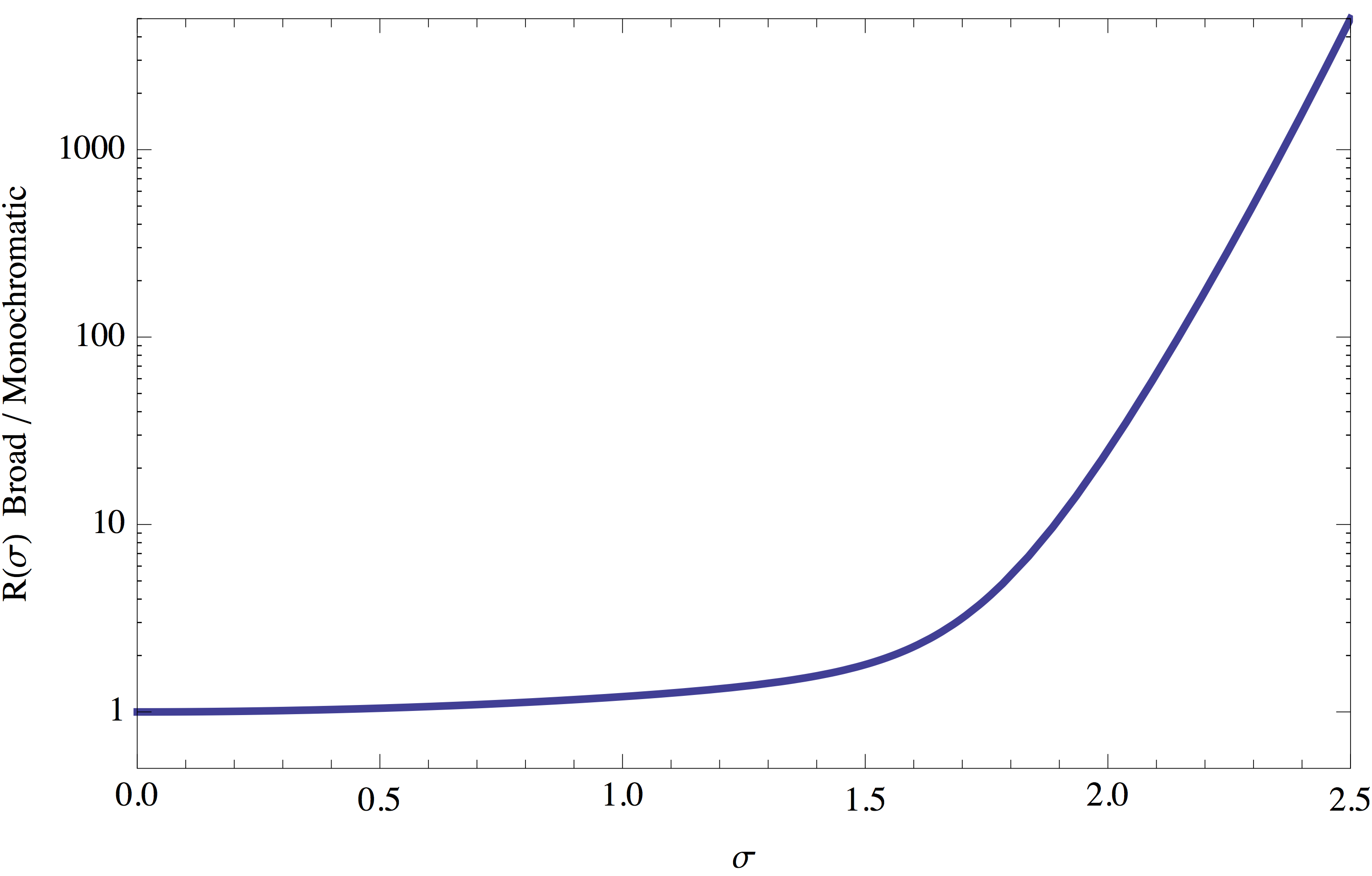}
    \vspace*{-5mm}
    \end{center}
   \caption{The function $R(\sigma)$ {rescaling the stochastic GW background for a broad PBH mass spectrum, compared to the monochromatic $(\sigma = 0)$ case.}  }
   % describes the deviation from one due to the width of the PBH mass distribution.} 
    \label{fig:ratio}
\end{figure}

We now focus on the case where all the PBH do not have the same mass but instead  follow a broad, lognormal, distribution 
\be
\rho(\mPBH)= \frac{\delta_{\rr{PBH}}^{\rr{loc.}} \rho_{\rm DM}^0 }{\sqrt{2 \pi \sigma_{\rr{\rm PBH}}^2} } \exp \left[ - \frac{\log_{10}^2 (\mPBH / \mu_{\rr{\rm PBH}})}{2 \sigma_{\rr{\rm PBH}}^2} \right]~.
\ee
of central mass $\mu_{\rr{\rm PBH}}$ and  width $\sigma_{\rr{\rm PBH}}$ (note that logarithms are $\log_{10}$). % in units of $\log \mPBH $, 
%Then the individual merging rate of PBH with masses $m_{\rr B}$ and $m_{\rr A}$ in the range $m_{\rm min} < m_{\rr A} < m_{\rr B}$ reads
%\be \label{eq:ind_rate_broad}
%\tau(m_{\rr B}) = \int_{m_{\rm min}}^{m_{\rr B}} \tau_{\rr{\rm PBH}}^{\rr{capt}} (m_{\rr B}, m_{\rr A})   \dd(\log m_{\rr A}) 
%\ee
%where $ \tau_{\rr{\rm PBH}}^{\rr{capt}} (m_{\rr B}, m_{\rr A})  $ is given by Eq.~(\ref{eq:capture_rate}).  
In that case, the total merging rate per unit volume and per logarithmic interval of masses, which is the one required in Eq.~(\ref{eq:hc2}), reads 
\ba \label{eq:tot_rate_broad}
&&\tau_{\rm merg} (m_{\rm A},\,m_{\rm B}) =   \frac{  \tau_{\rr{\rm PBH}}^{\rr{capt}} \, \rho(m_{\rm B})\, f_{\rm DM}}{m_{\rr B}\, \del } \\
&&\hspace{3mm}= f_{\rm DM}\,\del\  \OM^2 \ \frac{\rhoc^2}{\Msun^2}\
2 \pi c\, \left( \frac{85 \pi}{6 \sqrt 2} \right)^{2/7} \!
\left(\frac{c}{v_{\rm rel}}\right)^{11/7} \nonumber \\  
&&\hspace{5mm} \times  \left(\frac{G\Msun}{c^2}\right)^2 (m_{\rm A} m_{\rm B})^{2/7} (m_{\rm A} + m_{\rm B})^{10/7} \!
\frac{1}{2\pi\sigma^2} \nonumber \\  
&&\hspace{5mm} \times 
\exp\left[ - \frac{\log_{10}^2 (m_{\rm A} / \mu)}{2 \sigma^2} 
 - \frac{\log_{10}^2 (m_{\rm B} / \mu)}{2 \sigma^2}\right]\,.
\ea
The total rate is given by one half the integral of $\tau_{\rm merg} (m_A,m_B)$ over the whole mass range.  As in the monochromatic case, the local density contrast only rescales the merging rate linearly.   As discussed in~\cite{Clesse:2016vqa}, increasing the width of the mass distribution enhances the merging rate.  The amplitude (\ref{eq:hc2}) then becomes
\ba \label{eq:strain}
h_{\rr c}^2(f) &=& \frac{4}{3 \pi^{1/3}}\left(\frac{G\Msun}{c^2}\right)^{5/3}
\left(\frac{f}{c}\right)^{-4/3} \,A_1\times {\rm Int}\ \\
{\rm Int} &=& \frac{1}{\ln^{2}\!10} \int\!\!\int \frac{\dd m_{\rm A}}{m_{\rm A}} \,\frac{\dd m_{\rm B}}{m_{\rm B}} \,\tau_{\rm merg}(m_{\rm A},\,m_{\rm B}) \, \Mc^{5/3}
\nonumber \\[2mm]
&=& 2.16\times10^{-10}\ f_{\rm DM}\,\del\ {\rm yr}^{-1}{\rm Gpc}^{-3}
\times {\cal I} \nonumber
\ea
where
\ba  
{\cal I} &=& \mu^{5/3}\,e^{\frac{25}{36}\frac{\sigma^2}{\ln^2\!10}}\left[\sum_{n=0}^3
\frac{\alpha!}{n!(\alpha-n)!}e^{\left(n - \frac{\alpha}{2}\right)^2\frac{\sigma^2}{\ln^2\!10}} 
\right.   \\
&& \hspace{-3mm}\left.+\ \frac{1}{2}\,\frac{\alpha!}{4!(\alpha-4)!}
e^{\left(4 - \frac{\alpha}{2}\right)^2\frac{\sigma^2}{\ln^2\!10}}\right]  \nonumber
\equiv 2^\alpha\,\mu^{5/3} \times R(\sigma)\,,  \label{eq:Rsigma}
\ea
with $\alpha=23/21$. Note that for a monocromatic mass distribution peaked at $M=\mu$, we would have $R(\sigma=0)=1$.

For broad spectra with $\sigma\approx 2$ it was found that lower local density contrasts $10^6 \lesssim \delta_{\rr{PBH}}^{\rr{loc.}} \lesssim 10^{8}$ can produce merger rates in agreement with AdvLIGO limits.   In section~\ref{sec:GWbkg} we discuss how this can affect the induced stochastic background of gravitational waves. 
% {\bf JGB. I find a huge enhancement for this value of $\sigma$, see Fig.~3.} 
 
 \subsection{Extended halo mass function}  \label{sec:exthmf}

 {
The two previous models (monochromatic and broad mass distribution) assume a natural clustering scale, encoded in the parameter $\del$.   One can instead consider an extended halo mass function (and for simplicity a monochromatic PBH mass distribution) that would follow the expectations for the standard merger tree paradigm  in $\Lambda$-CDM.   Once the halo mass function is chosen, one can then apply Eq.~(\ref{eq:hc2}) to get the stochastic GW spectrum.   In practice, there are lot of uncertainties on the number density of halos of mass below $\sim 10^8 M_\odot$.   Nevertheless, an analytical estimation from the Press-Schechter formalism, as well as low-mass extrapolations of N-body simulations have been used by~\cite{Bird:2016dcv} to calculate the PBH merging rate in different halo models.  When integrated over halo density profiles and the halo mass function, the total merging rate is approximatively given by
\be
\tau_{\rm merg} \simeq 2.0  \times f_{\rm HMF}  \left(\frac{M_{\rm c}}{400 M_\odot}\right)^{-11/21}  \ \rr{yr}^{-1} \rr{Gpc}^{-3}, \label{eq:Birdrate}
\ee
where the parameter $f_{\rm HMF}$ depends on halo mass function ($f_{\rm MF} \simeq 1$ for a Tinker mass function, $f_{\rm MF} \simeq 0.6 $ for a mass function derived from the Press-Schechter formalism).  The critical halo mass  $ M_{\rm c}$ is the one below which halo evaporation takes place in a timescale less than about 3 Gyr.   In the case of PBHs with $30 M_\odot$ mass, one expects $M_{\rm c} \approx 400 M_\odot $~\cite{Bird:2016dcv} but this value is subject of large uncertainties, e.g. related to possible halo concentrations.   The halo mass function is the principal redshift dependent variable, so one needs in principle to include this dependance in the redshift integral of Eq.~(\ref{eq:hc2}).  Nevertheless, on one hand, simulations suggest that it is relatively stable over the redshift range $0<z<3$~\cite{Jenkins:2000bv,2013MNRAS.433.1230W}.  On the other hand, the integral is suppressed at high-redshift by $H(z)^{-1}(1+z)^{-4/3}$, so that the contribution of GW from redshifts $z\gtrsim 3$ is subdominant.  This also make the GW spectrum relatively insensitive to changes in PBH clustering and velocities at high redshifts.  For these reasons, and since the scope of the paper is not to compute GW spectra with a high accuracy, it is safe to consider in first approximation that the PBH merging rates are constant in time, and to plug Eq.~(\ref{eq:Birdrate}) in Eq.~(\ref{eq:hcf}) to get the expected amplitude of the stochastic GW background.  One thus gets
\be
h_{c} = 1.6 \times 10^{-25}  f_{\rm MF}^{1/2} \left(\frac{M_{\rm c}}{400 M_\odot}\right)^{-11/42} \left(\frac{f}{\rr{Hz}}\right)^{-2/3}\hspace{-1mm}\left( \frac{\Mc }{\Msun}\right)^{5/6}~.
\ee
In this scenario, the large majority of merging events occur within low-mass halos with mass $M_{\rm halo} \lesssim 10^5 M_\odot$, for which the typical Virial velocity $v_{\rm vir} \lesssim 1$ km/s.  As explained later, such a low value leads to the formation of PBH binaries with initially large semi-major axis, which avoid the suppression of GW of low frequencies due to limited orbital radius when PBH velocities are large.
}

\section{Detectability of the Stochastic Gravitational Wave Background}  \label{sec:GWbkg}

\begin{figure*} 
\begin{center}
\includegraphics[scale=0.9]{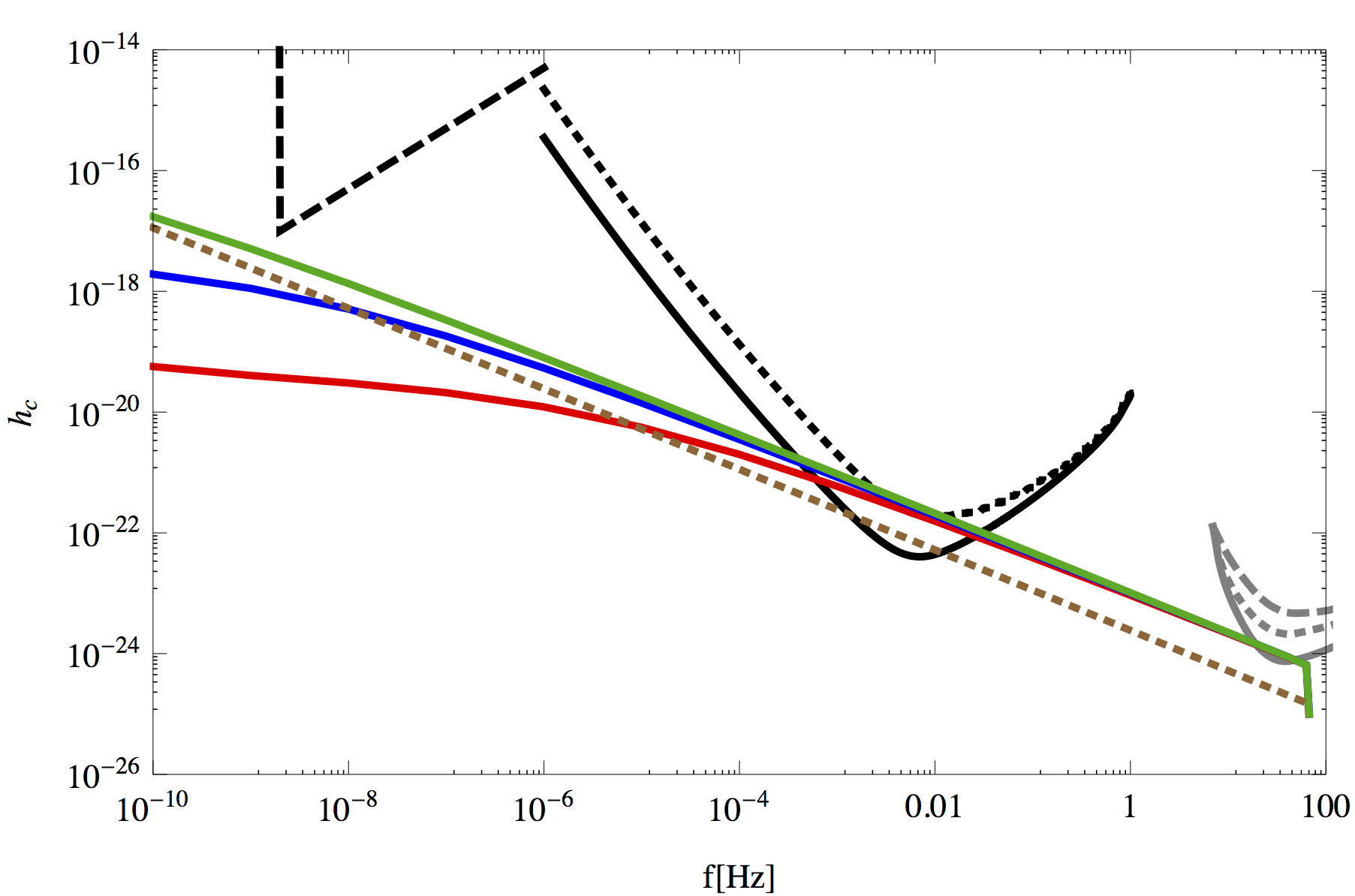}~\includegraphics[scale=0.9]{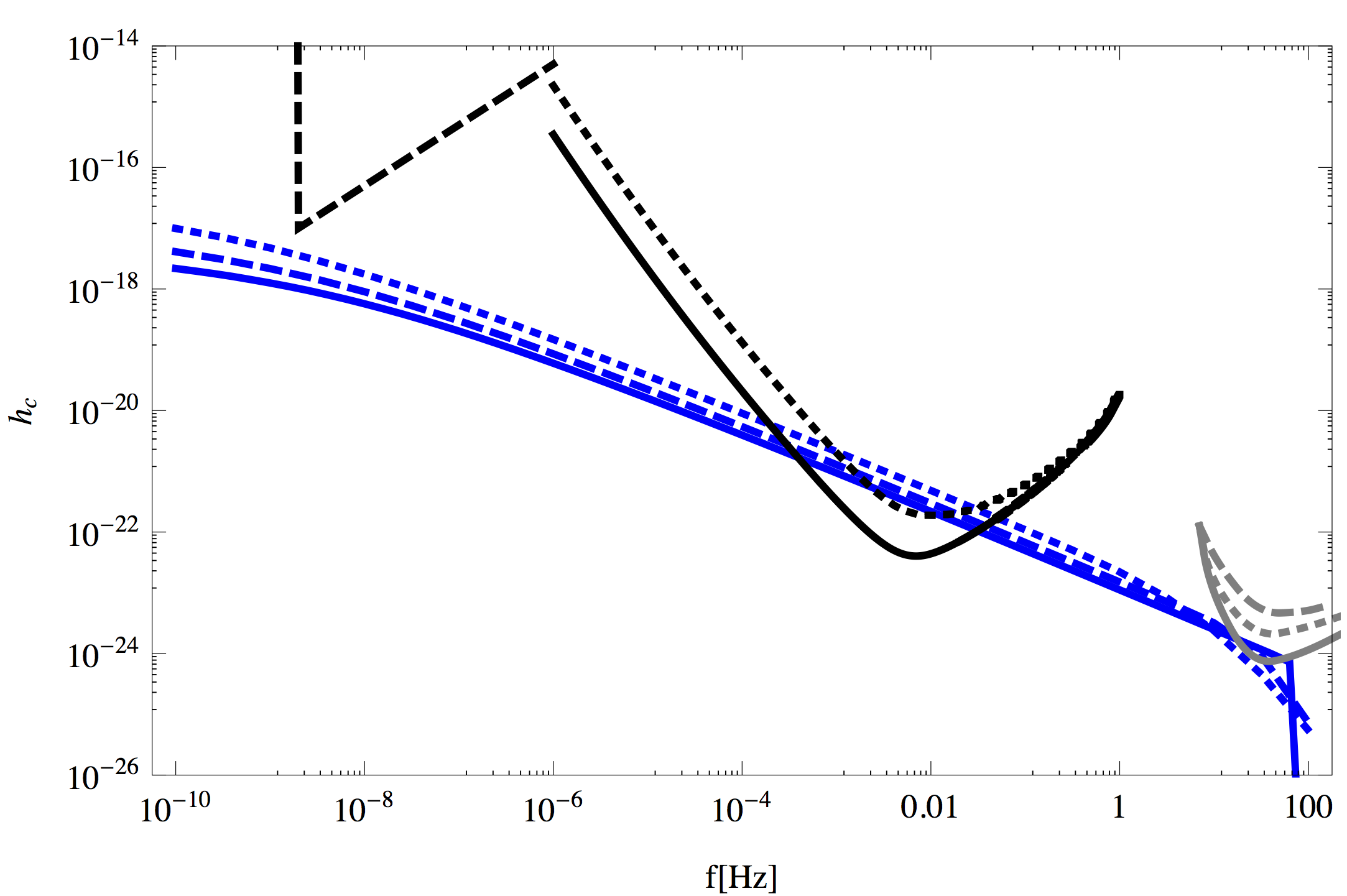} 
\includegraphics[scale=0.9]{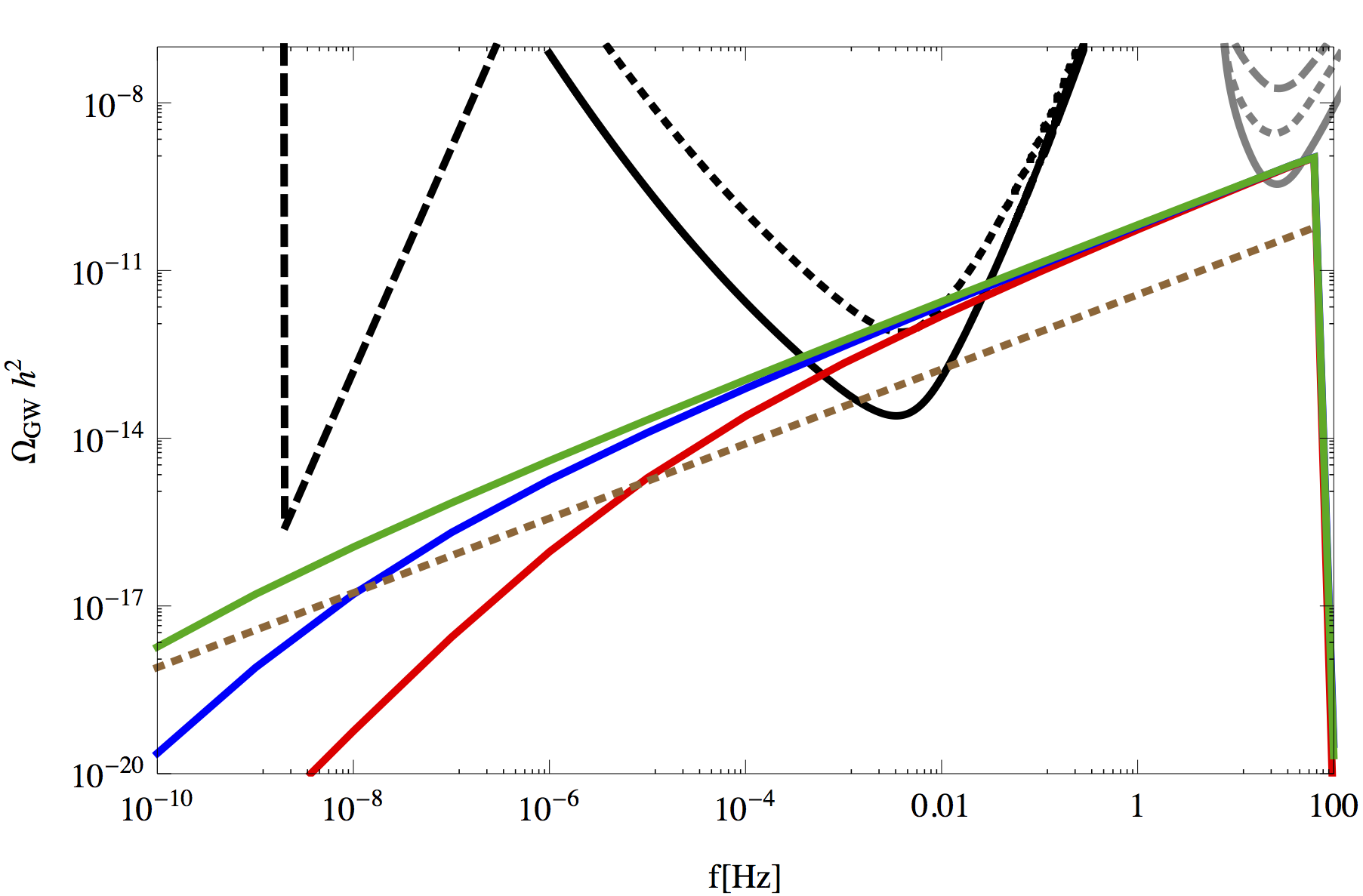}~\includegraphics[scale=0.9]{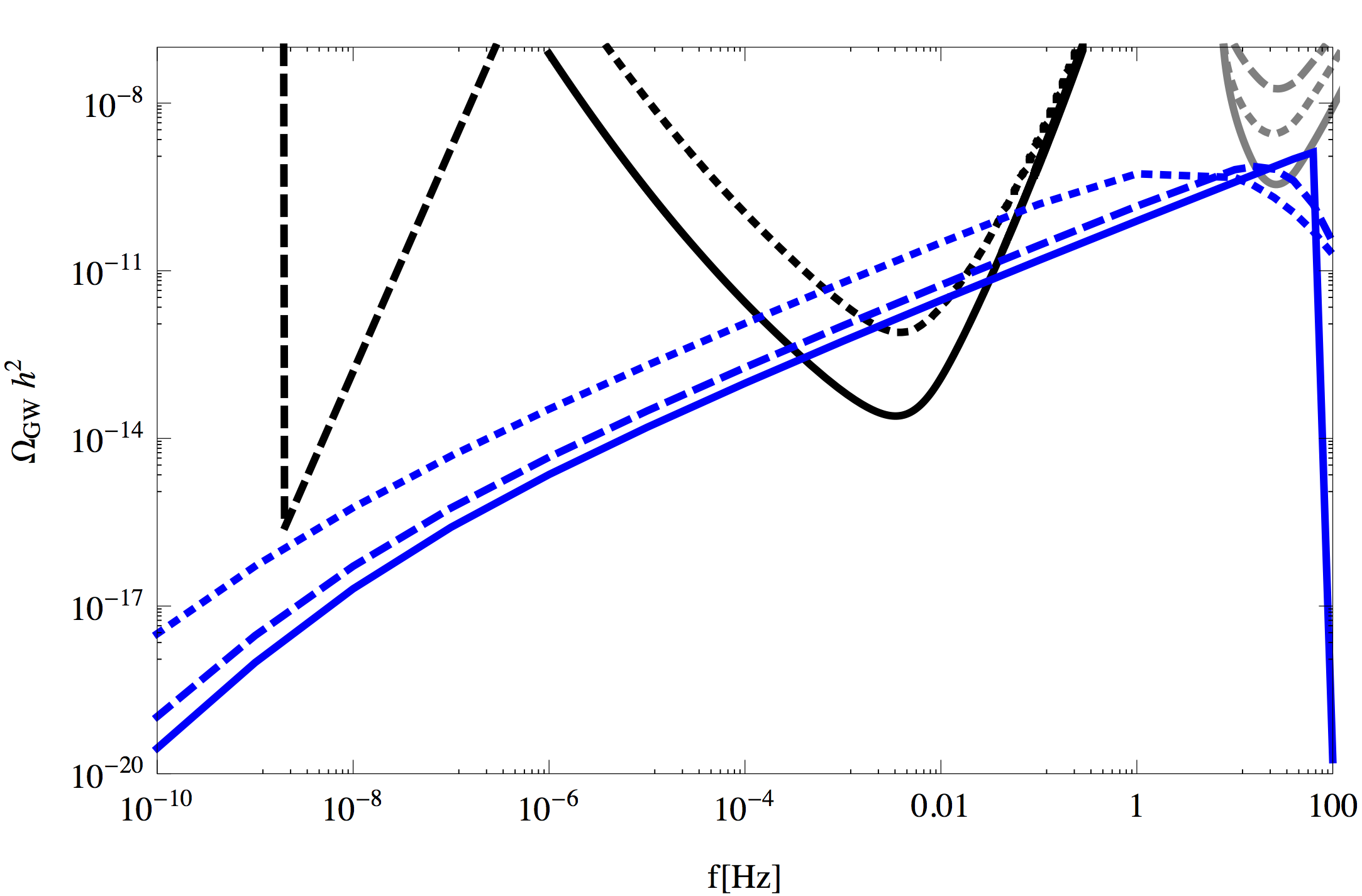}
\caption{\textit{Top panels.} Left: Stochastic GW spectrum $h_{\rr c}$ for a close to monochromatic ($\sigma = 0.1$) PBH mass distribution with $\mu = 30 \, \Msun$ and $v_{\rm vir} = 2 $ (green), $20$ (blue) and $200 \, \rr{km/s}$ (red).   Density contrasts are normalized to produce a constant merger rate $\tau = 50$ yr$^{-1}$ Gpc$^{-3}$, which corresponds respectively to $10^{-8} \delta_{\rm PBH}^{\rm loc}  = 0.2 / 8.0 / 290 $.   {The expected GW background for Bird et al. model~\cite{Bird:2016dcv} (extended halo mass function) is also represented (dotted brown)}.  Right:  GW spectrum for $\mu = 30 \Msun$ and  $v_{\rm vir} = 20 \, \rr{km/s}$ in the broad mass spectrum case, with $\sigma=  0.1 / 1 / 2$ (solid, dashed and dotted line respectively).  \textit{Bottom panels.}   The GW background $\Omega_{\rm GW} h^2$ for the same parameters.   Solid and Dotted black curves are the expected sensitivities for LISA respectively for the best and worse experimental designs.  Black dashed curve represents the sensitivity of the SKA through PTA observations.  The grey curves on the right represent the sensitivity of Advanced LIGO (O1 Run - dashed, O2 Run - dotted, O5 Run - solid). }
%\caption{Left: Spectrum of $h_{\rr c}$ (top) and $\Omega_{GW}$ (bottom) for a monochromatic mass distribution of PBH.  Solid lines are for PBH mass $\mPBH = 30 \Msun$ and merging rates $\tau = 10, 100, 1000$ yr$^{-1}$ Gpc$^{-3}$ (respectively blue, red and green lines).  Dotted and dashed red lines are for $\tau = 100$ yr$^{-1}$ Gpc$^{-3}$ and  $\mPBH = 10 \Msun$ and  $\mPBH = 100 \Msun$ respectively.  Right:  Spectrum of $h_{\rr c}$ (top) and $\Omega_{GW}$ (bottom) for a broad mass distribution of PBH, when varying $\sigma_{PBH}$.   Solid line is for  $\sigma_{PBH} = 0.1$, dashed line for $\sigma_{PBH} = 0.25$, and dotted line for $\sigma_{PBH} = 0.5$. }
\label{fig:GWspectra}
\end{center}
\end{figure*}

\begin{figure*}%[h!]
\begin{center}
\includegraphics[scale=0.45]{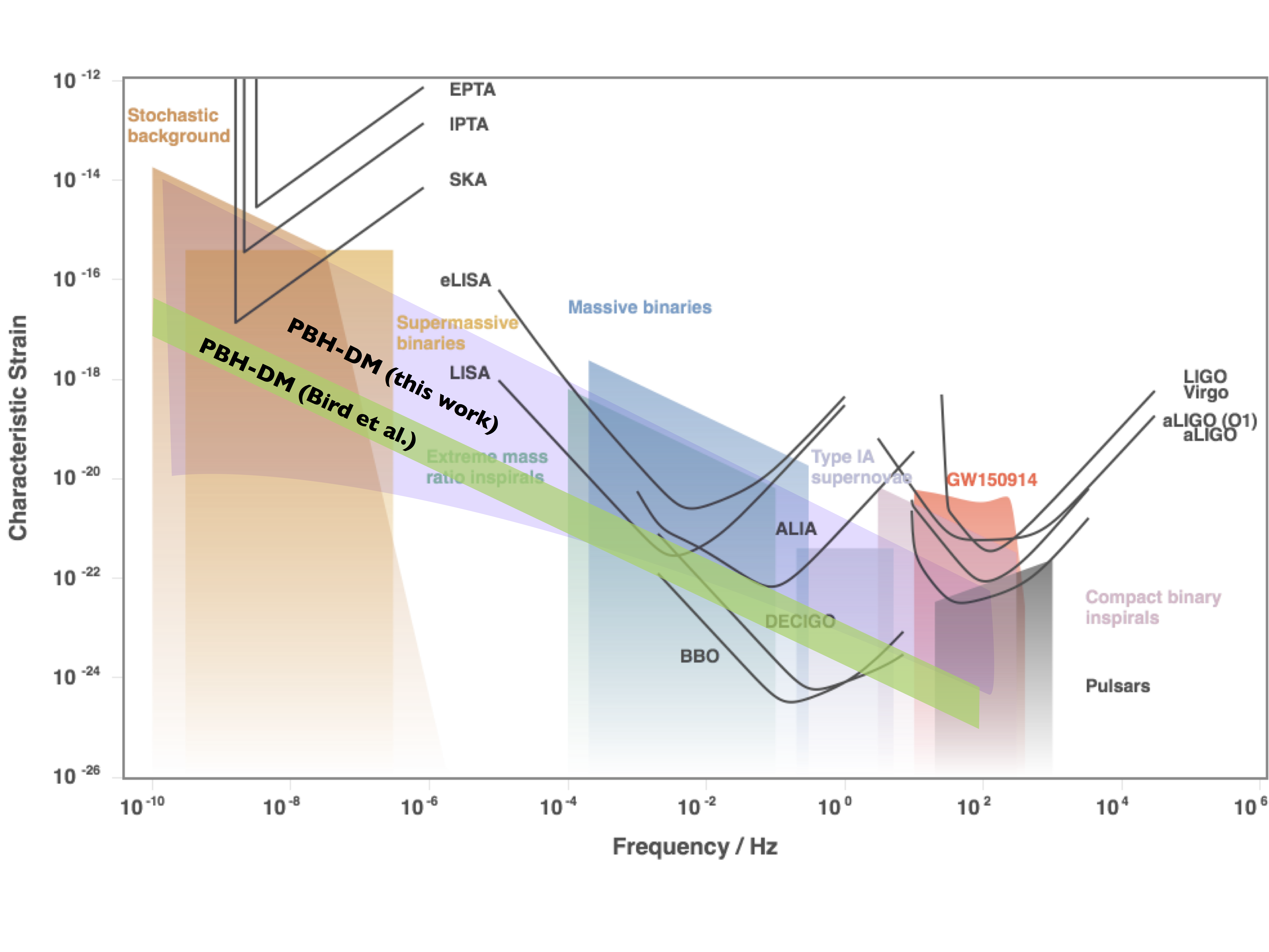}\\  \label{fig:strain_experiments}
\caption{Sketch displaying the expected limits for future GW experiments\protect\footnotemark
%\footnote{obtained using the \texttt{GWPlotter} tool~\cite{Moore:2014lga}, http://rhcole.com/apps/GWplotter/} 
 as well as the stochastic GW background for various astrophysical and cosmological processes.  The mauve band correspond to the expectations for the PBH-DM model considered in this paper, where PBH are regrouped in dense sub-halos, for merging rates consistent with the ones inferred by AdvLIGO, {i.e. between 8 and 240 events per year per Gpc$^3$, and mean PBH masses in the range $10_\odot \lesssim \mPBH \lesssim100 \Msun$.   Broad mass spectra allow to enhance the spectrum up to the present EPTA PTAs and LIGO limits.   For these merging rates, the stochastic GW background reaches the level of detectability of LISA and the SKA.  
 For comparison, the green band represents the region covered by the model of Bird et al.~\cite{Bird:2016dcv} extrapolated to lower frequencies. }      }
\end{center} 
\end{figure*}

In this section we compute the stochastic GW spectrum and explore the effects of the PBH mass distribution and of the clustering through the halo Virial velocity.  The marginalized GW spectrum over PBH masses (in the broad spectrum case), velocities and impact parameter $b$, have been computed from the posterior PDF of $h_{\rr c}$ obtained for a synthesized  population of $10^6$ binaries.   The two cases of a close to monochromatic and a broad mass spectrum are considered.   In each case, the detectability of the signal by space GW experiments and PTAs is discussed.    

If the limit of low Virial velocities ($v_{\rm vir} \lesssim 10$ km/s), corresponding to low-mass or extended PBH halos, the analytical estimation of the strain amplitude is valid for any width and mean of the PBH  mass distribution if the factor $R(\sigma) $ calculated in the previous section is included.   Analytical estimates are also valid at large frequencies (but lower than $f_{\rm max}$), for larger PBH velocities.   After combining Eqs.~(\ref{eq:strain}),~(\ref{eq:Rsigma}) and~(\ref{eq:merging}), the squared strain is given by
\ba
h_{\rr c}^2(f) &=& (1.14\times10^{-25})^2 \ f_{\rm DM}\,\del \ R(\sigma) \times \nonumber \\
&& 2.45\times10^{-9}\,
\left(\frac{\mu}{\Msun}\right)^{5/3}\left(\frac{f}{\rm Hz}\right)^{-4/3}\,.
\ea
One can then compute the stochastic GWB amplitude
\ba \nonumber
h^2\Omega_{\rm GW}(f) &=& 2\times10^{-23}\ f_{\rm DM}\,\del \times \\
&&\left(\frac{f}{\rm Hz}\right)^{2/3}\!\left(\frac{\mu}{\Msun}\right)^{5/3} R(\sigma)\,.
\ea

The numerical spectra are represented on Fig.~\ref{fig:GWspectra} for various values of $v_{\rm vir} $ and $\sigma$ and are consistent with analytical estimations in the low velocity/large frequency limit, in both cases of a (close to) monochromatic and broad mass distribution.  We also show the expected sensitivity of LISA for its worse and best designs, as well as the SKA sensitivity to GW through the observations of PTA, and the sensitivity of Advanced LIGO/Virgo (Runs O1, O2 and O5).

Even in the small-width limit of the mass spectrum, i.e. when $R(\sigma) \approx 1$, the GW background reaches the sensitivity of LISA for $\delta_{\rm PBH}^{\rm loc} \sim 10^{9}$, the exact value depending on the considered experimental design.   Therefore LISA will probe models producing merging rates consistent with Advanced LIGO, for instance models in which a non-negligible fraction of PBH-DM is clustered within ultra-faint dwarf galaxies.  

Moreover, given that $R(\sigma) $ blows up for $\sigma \gtrsim 2$, see Fig.~\ref{fig:ratio}, the LISA constraint on $  \delta_{\rm PBH}^{\rm loc} $ will be much more stringent in the broad spectrum case where PBH masses cover more than two decades, as in the model proposed in Ref.~\cite{Clesse:2015wea}.     As a consequence, in case of a signal detection, and assuming that Earth-based experiments will have somehow improved the measurements of BH merging rates, LISA will have the ability to determine how broad is the PBH mass spectrum~\cite{Bartolo:2016ami}.   

Using PTAs, the present limits on the strain $h_{\rr c} < 1.0 \times 10^{-15}$ (95\% C.L.) at a {frequency of $ 4\times 10^{-9}$Hz} from the Parkes radio-telescope~\cite{Shannon:2015ect} already exclude some models consistent with AdvLIGO rates (with negligible velocity effects), with PBH mean masses larger than $\sim 10$ stellar masses and broad mass spectra $\sigma \gtrsim 3 $.   With such a broad spectrum, the value of $f_{\rm max}$ is suppressed for a large fraction of the PBH binaries, and therefore the spectrum at frequencies probed by AdvLIGO is also suppressed, thus explaining why such models should not have yet been detected.  

 Numerical simulations allow to study the effect of PBH velocities on the GW spectrum at PTA and LISA frequencies, through the minimal frequency $f_{\rm min}$.   This minimal frequency is linked to the maximal impact parameter $b_{\rm max}$ in the two-body capture process, which is affected by the relative PBH velocity.   Increasing the velocity $w$ reduces  $b_{\rm max}$ and enhances $f_{\rm min}$ at such a point that an important fraction of the binaries cannot emit at PTA frequencies, and even a small but non-negligible fraction of them at larger frequencies in the range of LISA.   As a consequence, one observes that the characteristic  $ h_{\rr c} \propto  f^{-2/3}$ frequency dependence of a population of massive binaries is modified for velocities $v_{\rm vir}\gtrsim 20 \rm{km/s}$.   The spectrum is found to be strongly suppressed on PTA scales, and the slope of the spectrum is slightly modified on LISA frequencies.  This signature in the GW spectrum will be detectable by the best LISA design in the case of $\tau^{\rr{merg}} = 50 \, \rr{yr}^{-1}\rr{Gpc}^{-3}$ and $\mu = 30 \Msun$, i.e. the most relevant models after AdvLIGO results.   But even for less favorable LISA designs, the effect could still be detected for larger merging rates or larger mean PBH mass.   Finally one notes that the GW spectrum could also reach the limit of the observing Run O5 of Advanced LIGO/VIRGO.   
 
 { The expected GW background for the extended halo mass model proposed by Bird et al.~\cite{Bird:2016dcv} and discussed in Sec.~\ref{sec:exthmf} has also been represented on Fig.~\ref{fig:GWspectra}.  The GW background amplitude is lower, due to reduced merging rates, still within the range of detectability of the best LISA design but not of LIGO-O5.   The above mentioned velocity effect is naturally suppressed compared to models where PBHs are clustered in faint dwarf satellites, because most mergers occur in low-mass halos for which the expected PBH velocity is typically lower than a km/s.    }
 
 {The stochastic GW backgrounds shown in Fig.~\ref{fig:GWspectra} consider the $95\%$ C.L. emission from unresolved events at large distances.   However, there will possibly be 4 or 5 sigma detections by LISA of individual events which have much larger amplitude and will stand above the stochastic background.  But such events may not be distinguished from other SMBH events, and only the shape of the stochastic background may allow us to distinguish them. }
 
 Some other plausible sources of GW backgrounds and their possible range is shown on Fig.~\ref{fig:strain_experiments}.   A PBH stochastic background could be distinguished from these sources in two ways:  first, by using the specific frequency dependence that is found if PBH are somehow clustered and have a broad mass distribution; second, because PBH cover the whole range of frequencies from $10^{-10}$ to $100$ Hz.  The use of multi-frequency detectors and multi-probe analysis will thus be crucial to make this distinction. 

To summarize, PTA, space and Earth-based interferometers are very complementary probes of PBH-DM models.  Combining these probes will allow to constrain not only the PBH abundances but also their mass spectrum and their clustering properties.   They will also have the ability to distinguish between  primordial and stellar origins of the BH binaries through a specific frequency dependance of the GW spectrum.  {One can remark that a population of individual stellar black holes within dense star clusters should produce a similar frequency dependance.    In this case, we nevertheless believe that the BH abundance is highly suppressed compared to PBH dark matter abundances, and so is the stochastic GW background amplitude.   Detecting black holes with masses larger than $\mathcal O(100) M_\odot $ or lower than one $M_\odot$ will be the ultimate way to distinguish between stellar and primordial origins.  }

%\section{Detectability}

\footnotetext{obtained using the \texttt{GWPlotter} tool~\cite{Moore:2014lga}, \url{http://rhcole.com/apps/GWplotter/}}

\section{Discussion and conclusions} \label{sec:ccl}

The first detection of gravitational waves from massive inspiralling black holes by AdvLIGO has opened the new era of Gravitational Wave Astronomy. In the near future we will have an array of GW observatories or telescopes operating both on the ground and in space, with increasing precision and resolution. It is therefore essential to characterize and predict all the possible irreducible stochastic backgrounds that will limit our ability to detect individual events far away in the universe. For this purpose we have studied in this paper the expected irreducible GW background form the inspiralling of a broad distribution of black holes that could have started merging soon after recombination, before the reionization of the universe.

We have studied the evolution and emission of a primordial black hole population from the initial capture down to final merging, and found that the amplitude of this stochastic background can be significantly larger than previously estimated, due to the fact that the PBH are distributed in mass far from a monochromatic distribution. We took into account effects like the initial eccentricity of the PBH binaries, their initial velocities, mass distribution and clustering.   

We find that the stochastic background could {barely} be seen by AdvLIGO, nor by PTA, but be easily detectable by the future LISA space interferometer, if the PBH mass spectrum is close to monochromatic.  { Its amplitude is nevertheless strongly amplified and potentially detectable by all these experiments if the mass distribution is wide. } In fact, this background has become the most relevant irreducible background of gravitational waves for LISA~\cite{Bartolo:2016ami}. Therefore, it is extremely important to predict its characteristics with high accuracy. In particular, there are features of this background that had passed unnoticed in the past, like the slight turn over at low frequencies due to the limited range of initial semimajor axis allowed by the capture scenario we describe here. Moreover, we find that the initial eccentricity of the orbit will not play a significant role since gravitational wave emission makes all those highly eccentric orbits decay in less than a single orbital period for most of the parameter range.  

Finally, the fact that the PBH mass distribution is not monochromatic induces a significant increase in the amplitude of the stochastic background, for relatively moderate widths of the lognormal mass distribution, well above that predicted by Ref.~\cite{Cholis:2016xvo}, as illustrated on Fig.~\ref{fig:strain_experiments}. This increase comes from the merging of relatively low mass binaries with a large contribution to the PBH population. This feature will allow us to distinguish this background from the usual non-primordial binary black hole background from astrophysical origin, since no black holes of masses below a solar mass are possible in those scenarios.   Compared to Refs.~\cite{Mandic:2016lcn,Cholis:2016xvo}, we have included three new ingredients, a broad mass distribution, the effect of PBH clustering and we have considered the whole frequency range interesting for future GW detectors.  

{The GW spectra presented in the paper are obtained for a model with a constant merging rate, but all the above mentioned effects are generic for any PBH model, and our results can easily be extended to models with time or space dependent merging rates. } {Different clustering scenarios for the low-mass halos could also be considered and would need specific N-body simulations.  But one should keep in mind that PBHs do  \textit{not require} sizable modifications of the standard merger tree paradigm (see e.g.~\cite{Knebe:2001en} for the effects of primordial power spectrum enhancement).}

In the future, we will be able to test the parameters of the inflationary model responsible for the generation of the peak in the spectrum of density fluctuations, a few e-folds before the end of inflation, that could give rise to these PBH, and thus open a window into the physics of the early universe, well above that accessible to high energy particle physics accelerators.

\section*{Acknowledgments}  

We thank Marco Peloso, Caner Unal, Jens Chluba, Yacine Ali-Ha\"imoud, Alvise Raccanelli and the LISA Cosmology WG for useful discussions.
This work is supported by the Research Project of the Spanish MINECO, FPA2013-47986-03-3P, and the Centro de Excelencia Severo Ochoa Program SEV-2012-0249. 

\bibliography{biblio}

\end{document}